\documentclass[aps,prl,superscriptaddress,twocolumn,floatfix,a4paper]{revtex4}

\usepackage{graphicx,graphics,epsfig}   
\usepackage{dcolumn}    
\usepackage{bm}         
\usepackage{amsmath,amssymb,amsthm,amstext}    
\usepackage{verbatim}   
\usepackage{color}      
\usepackage{subfigure}  
\usepackage{times,natbib}
\usepackage{amsmath,amsfonts,amssymb,graphics,graphics,color,times}

\usepackage{latexsym}
\usepackage{amsmath}
\usepackage{amssymb}
\usepackage{amsfonts}
\usepackage{amsthm}
\usepackage{mathrsfs}
\usepackage{color,verbatim,graphics}
\usepackage{psfrag}
\DeclareMathAlphabet{\mathrsfs}{U}{rsfs}{m}{n}
\DeclareMathAlphabet{\mathpzc}{OT1}{pzc}{m}{it}
\DeclareMathAlphabet{\matheus}{U}{eus}{m}{n}
\DeclareMathAlphabet{\mathbbold}{U}{bbold}{m}{n}

\def\be{\begin{equation}}
\def\ee{\end{equation}}
\def\bea{\begin{eqnarray}}
\def\eea{\end{eqnarray}}
\newcommand{\ban}{\begin{eqnarray*}}
\newcommand{\ean}{\end{eqnarray*}}

\newcommand{\ket}[1]{|#1\rangle}
\newcommand{\bra}[1]{\langle#1|}
\newcommand{\braket}[2]{\langle#1|#2\rangle}

\definecolor{nblue}{rgb}{0.2,0.2,0.7}

\begin{document}

\title{Dimension witnesses and quantum state discrimination}

\author{Nicolas Brunner}
\affiliation{D\'epartement de Physique Th\'eorique, Universit\'e de Gen\`eve, 1211 Gen\`eve, Switzerland}
\affiliation{H.H. Wills Physics Laboratory, University of Bristol, Tyndall Avenue, Bristol, BS8 1TL, United Kingdom}
\author{Miguel Navascu\'es}
\affiliation{H.H. Wills Physics Laboratory, University of Bristol, Tyndall Avenue, Bristol, BS8 1TL, United Kingdom}
\author{Tam\'as V\'ertesi}
\affiliation{Institute of Nuclear Research of the Hungarian Academy of Sciences
H-4001 Debrecen, P.O. Box 51, Hungary}



\begin{abstract}
Dimension witnesses allow one to test the dimension of an unknown physical system in a device-independent manner, that is, without placing assumptions about the functioning of the devices used in the experiment.  
Here we present simple and general dimension witnesses for quantum systems of arbitrary Hilbert space dimension.
Our approach is deeply connected to the problem of quantum state discrimination, hence establishing a strong link between these two research topics. Finally, our dimension witnesses can distinguish between classical and quantum systems of the same dimension, making them potentially useful for quantum information processing.
\end{abstract}

\maketitle

Recently, the problem of testing the dimension of an unknown physical system has attracted quite some attention.
Here dimension 
represents, loosely speaking,
the number of degrees of freedom of the system. For quantum systems this corresponds to the Hilbert space dimension.
The main point of this line of research is to assess the dimension of an unknown physical system in a 'device-independent' manner, that is from measurement data alone, without any a priori assumption about the devices used in the experiment.

This is in contrast with the more usual approach in physics, in which, when constructing a theoretical model aiming at explaining some experimental data, the dimension of the system is a parameter that is defined a priori. For instance, when describing quantum systems, one generally starts by fixing the Hilbert space dimension, given reasonable assumptions about the nature of the system and its dynamics.
Then the model may or may not reproduce the experimental data. If the model fits the data, one can make a statement about the system's dimension. If the model does however not work, nothing can be said since, in principle, there could be a different model using the same dimension that could explain the data. Obviously testing all possible models with a fixed dimension is impossible hence a better approach is required in order to determine the minimal dimension of the system compatible with some data.

The concept of a dimension witness was recently introduced to address this problem. First discussed in the context of Bell inequalities \cite{brunner,pal,peresgarcia,briet,junge2}, dimension witnesses were also derived in the case of quantum random access codes \cite{wehner} and the time evolution of quantum observables \cite{wolf}.

More recently a framework was developed in order to derive dimension witnesses in a 'prepare-and-measure' scenario \cite{gallego}, which represents the simplest but also the most general case. Consider the experiment of Fig.~1. A first device prepares on demand an unknown physical system in one out of $N$ possible states $\rho_x$. A second device then performs one out of $m$ possible measurements, giving an outcome $b=1,...,k$. The experiment is described by a set of conditional probabilities $P(b|x,y)$, which represent the probability of observing outcome $b$ when state $x$ was prepared and measurement $y$ was performed.
In each round of the experiment, which state $x$ is prepared and which measurement $y$ is performed is chosen by the observer. However, the important point is that, what the states and measurements actually are, is unknown to the observer. The observer's task will then be to estimate the minimal dimension $d$ of the system that is compatible with the experiment. More precisely what is the minimal dimension necessary to reproduce a given set of conditional probabilities $P(b|x,y)$?

 \begin{figure}[b!]
  \includegraphics[width=0.7\columnwidth]{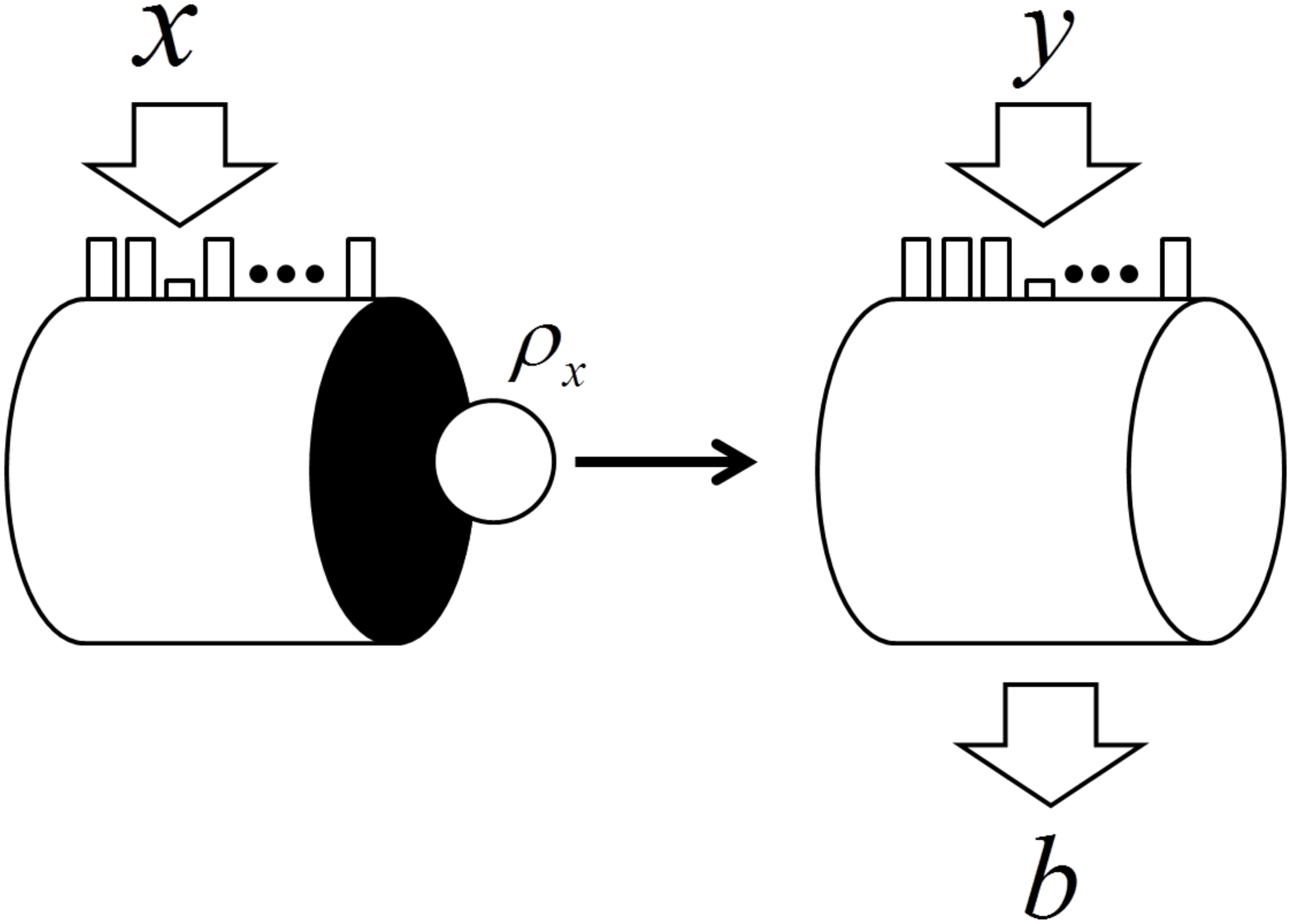}
  \caption{Testing the dimension of an unknown system in a prepare-and-measure scenario.}
\label{fig}
\end{figure}

In the case of classical systems, the above problem can be tackled using the framework of Ref.~\cite{gallego}, which allows one to derive dimension witnesses for classical systems, based on geometrical ideas.
For quantum systems however, 
finding dimension witnesses for systems of arbitrary Hilbert space dimension is challenging, and no general solution has been provided yet.
The approach of Ref.~\cite{gallego} can be used to derive quantum dimension witnesses, but the validity of these can generally be tested only numerically. Hence this approach can be used only for systems of relatively low dimension. The approach of Ref.~\cite{wehner} does not rely on numerics, but gives witnesses for which strong and/or tight bounds can be derived only in particular cases. Finally, the approach of Ref.~\cite{wolf} does not apply to the static case.

The problem of testing the dimension of quantum systems in a prepare and measure scenario is well motivated, and not only from a purely conceptual point of view.
Indeed, in quantum information, the dimensionality of quantum systems represents a resource for information processing. In general, systems of larger dimension offer more power for computation and communication. In particular, they are known to simplify quantum logic \cite{lanyon}, enable for the optimal implementation of certain quantum protocols \cite{rudolph}, and allow for lower detection efficiencies in Bell tests \cite{massar,vertesi4}. Moreover, the dimension of quantum systems plays a crucial role in the security of standard quantum cryptography \cite{acingisin}. Finally, dimension witnesses are relevant in practice, as demonstrated by two recent experiments \cite{hendrych,ahrens}, and several novel quantum information protocols are based on dimension witnesses, e.g. for quantum cryptography \cite{marcin} and randomness expansion \cite{hongwei}.

In the present paper, we present a general and simple method for deriving dimension witnesses for quantum systems of arbitrary dimension. 
Our approach is based on the distinguishability of quantum states, that is how well two (or more) quantum states can be distinguished from each other by performing the most general measurement. Thus our work
highlights a strong link between quantum dimension witnesses and the problem of quantum state discrimination (see e.g. \cite{helstrom,holevo,chefles,croke}), which has received considerable attention in the last years.

We start by introducing the basic notions and notations.
An experiment as described in Fig.~1 admits a quantum $d$-dimensional representation when
\be P(b|x,y) = \text{tr}(\rho_x M^b_y ) \ee
for some states $\rho_x$ and measurement operators $M^b_y$ acting on $\mathbb{C}^d$.
We also say that the experiment admits a classical $d$-dimensional representation when all states in the set $\{\rho_x \}$ commute pairwise.
In this case, the states can be represented as classical states of dimension $d$, i.e. probability distributions over dits.

Since characterizing the set of experiments that admit a quantum or classical $d$-dimensional representation is in general a difficult problem, it is convenient to introduce dimension witnesses. Consider a function that associates to each probability distribution $P(b|x,y)$ a real number. A function $f$ is then termed a quantum dimension witness if
\be\label{dimwit} f[P(b|x,y)] \leq Q_d \ee
for all experiments involving quantum systems of Hilbert space dimension smaller or equal to $d$, and there exists a set of data $P^*(b|x,y)$ involving systems in higher dimensions such that $f[P^*(b|x,y)]>Q_d$. Classical dimension witnesses are defined in a similar way; in this case the upper bound in
\eqref{dimwit} is denoted $C_d$.

Before we proceed with the presentation of our main results, it is instructive
to understand why the distinguishability of states is important here.
Consider first the case $d \geq N$. Here, the dimension $d$ of the system, the mediating particle, is in fact large enough to encode perfectly the choice of preparation $x$. Hence, it is then possible that the state preparator simply sends $x$ to the measuring device. The latter then has all information about both $x$ and $y$, and can thus simulate any statistics $P(b|x,y)$. Thus, if by measuring the mediating particle it is possible to perfectly identify which preparation $x$ was chosen, no relevant device-independent statement can be made about the dimension of the system.

On the other hand, when $d<N$, the choice of preparation $x$ cannot be encoded perfectly in the system anymore, since the latter cannot be prepared in sufficiently many perfectly distinguishable states. Therefore, not all statistics $P(b|x,y)$ can be reproduced when $d<N$, and relevant device-independent statements can be made.

Thus, a central aspect of the problem is how good one can distinguish between a set of $N$ quantum states. Intuitively, a good strategy consists in choosing the states to be as far from each other as possible in the Hilbert space, in order to make them as distinguishable as possible.
However, the dimension of the Hilbert space in which the set of states is embedded will clearly put restrictions on their distinguishability. It is exactly this aspect that we will exploit to derive our dimension witnesses.

The distinguishability of quantum states is captured by the notion of trace distance. Given two quantum states $\rho_x$ and $\rho_{x'}$, how good they can be distinguished from each other, allowing for the most general measurement, is quantified by the trace distance
\be D(\rho_x,\rho_{x'}) = \frac{1}{2} || \rho_x - \rho_{x'} ||_1. \ee
Operationally, the trace distance represents the maximal distance in probabilities that a POVM element $M$ may occur depending if the state is $\rho_x$ or $\rho_{x'}$, that is
\be \label{rel1} D(\rho_x,\rho_{x'}) = \max_{M} \text{tr} (  (\rho_x - \rho_{x'})M). \ee
Also, the trace distance can be related to another notion of distinguishability, the fidelity $F$ between $\rho_x$ and $\rho_{x'}$:
\be \label{rel2} 1-F(\rho_x,\rho_{x'}) \leq D(\rho_x,\rho_{x'}) \leq \sqrt{1-F^2(\rho_x,\rho_{x'})} \ee
Indeed if both states are pure, $F(\rho_x,\rho_{x'}) = | \braket{\Psi_x}{\Psi_{x'}}| $, and the second inequality in \eqref{rel2} is saturated.

Consider the scenario of Fig.~1. Take the simple case in which there are $N$ possible preparations, and a single measurement (i.e. $m=1$) with $N$ outcomes. In this case, we can build a dimension witness based on the average guessing probability, i.e. the function

\be U_N = \frac{1}{N}\sum_{x=1}^N P(b=x|x). \ee
In order to show that $U_N$ works as a quantum dimension witness for any $d<N$, we must find an upper bound on $U_N$ depending on $d$.
This can be done as follows

\be U_N = \frac{1}{N} \sum_{x=1}^N \text{tr}(\rho_x M_x) \leq \frac{1}{N}\sum_x\text{tr}(M_x)  =  \frac{d}{N} = Q_d \ee
hence leading to a dimension witness for any $d<N$.
Note however, that for this witness we have that $C_d=Q_d$ (for any $d\leq N$), hence the witness cannot distinguish between classical and quantum states of the same dimension.

After this warm-up, let us now see how to construct dimension witnesses for systems of arbitrary dimension $d$ such that $C_d<Q_d$. Consider again $N$ possible preparations, but now $m= \left( \begin{array}{c} N \\ 2 \end{array} \right)
=N(N-1)/2$ dicotomic measurements (with outcomes $\pm1$), labelled as $y=(x,x')$, with $x,x'\in\{1,...,N\},x>x'$.
Consider the following expression
\be
W_N\equiv \sum_{x>x'}|P(x,(x,x'))-P(x',(x,x'))|^2
\ee
where we used the simplified notation $P(x,(x,x')) \equiv P(b=1|x,(x,x'))$.

We will now see how to upper bound $W_N$ depending on $d$.
For each measurement $y=(x,x')$, call $M_{(x,x')}$ the POVM element corresponding to outcome $b=1$. From the structure of $W_N$, it is clear that each $M_{(x,x')}$ can be taken to be the projector onto the subspace generated by the positive eigenvectors of $\rho_x-\rho_{x'}$. Also, since $W_N$ is a convex functional, it follows that, in order to compute its maximum value, we can assume each of the states $\{\rho_x\}$ to be pure, i.e., $\rho_x=\ket{\Psi_x}\bra{\Psi_x}$. We thus have that

\bea
W_N&=& \sum_{x>x'} | \text{tr}
\{ (\rho_x-\rho_{x'})M_{(x,x')}\}|^2  \\  \nonumber
&\leq&\sum_{x>x'} | D(\rho_x,\rho_{x'}) |^2 
\leq \sum_{x>x'}(1-|\braket{\Psi_x}{\Psi_{x'}}|^2)
\eea
where we have used relations \eqref{rel1} and \eqref{rel2}.
Next we use the fact that

\bea
\sum_{x>x'}|\braket{\Psi_x}{\Psi_{x'}}|^2 &=&   \frac{1}{2}  \bigg[ \sum_{x,x'}|\braket{\Psi_x}{\Psi_{x'}}|^2 -N \bigg]  \\
 &= &  \frac{N^2}{2}\text{tr}(\Omega^2) - \frac{N}{2}
\label{interm}
\eea
with $\Omega=\frac{1}{N}\sum_{x=1}^N \ket{\Psi_x}\bra{\Psi_x}$ being a normalized quantum state. Since the purity of any $d$-dimensional normalized state $\Omega$ is lower bounded by
\be  \text{tr}(\Omega^2) \geq \frac{1}{d} \ee
we obtain that

\bea \label{QuadWit}
W_N &\equiv & \sum_{x>x'}|P(x,(x,x'))-P(x',(x,x'))|^2 \nonumber \\
&\leq& Q_d=\frac{N^2}{2}\left(1-\frac{1}{\min(d,N)}\right).
\eea
Thus $W_N\leq Q_d$ is a quadratic quantum dimension witness for any $d<N$.

\begin{table}
	\begin{ruledtabular}
	\begin{tabular}{c||cccccc}
	d &
	2 &
	3 &
	4 &
	5 &
	6 &
	7  \\
	\colrule
	$C_d$&  12   & 16 &   18 &   19 &   20 &   21\\
	$Q_d$& 12.25 &   16.33 &   18.38 &   19.60 &   20.42 &   21 \\
	\end{tabular}
	\end{ruledtabular}
\caption{\label{tab:table1} Tight bounds for $d$-dimensional classical ($C_d$) and quantum ($Q_d$) systems for the dimension witness $W_7$. Notably, the witness can distinguish between classical and quantum systems of the same dimension for any $d<N$, since $C_d<Q_d$.}
\end{table}

An interesting feature of the above witness is its tightness. That is, for any dimension $d$, there exists an ensemble of states $\{\rho_x\}_{x=1}^N\subset B({\mathbb C}^d)$ and measurement operators $\{S_y\}\subset B({\mathbb C}^d)$ which saturate the inequality \eqref{QuadWit}. Suppose that, for any $d \leq N$, there exists a set of pure states $\{\ket{\Psi_x}\}_{x=1}^N\subset B({\mathbb C}^d)$ such that $\Omega=\frac{1}{N}\sum_{x=1}^N \ket{\Psi_x}\bra{\Psi_x}=\frac{1}{d} \openone_d$. Then $\text{tr}(\Omega^2)=\frac{1}{d}$. Thus, by choosing $M_{(x,x')}$ to be the measurement that optimally discriminates between $\ket{\Psi_x}$ and $\ket{\Psi_{x'}}$, all inequalities incurred into in the derivation of the bound will turn into equalities, therefore proving the attainability of the bound. Hence it suffices to prove that such vectors exist. It can be verified that the normalized vectors

\be
\ket{\Psi_x}=\frac{1}{\sqrt{d}}\sum_{k=0}^{d-1}e^{\frac{i2\pi kx}{N}}\ket{k}
\label{vectors}
\ee
do the trick.

Next, we can derive a tight upper bound for classical systems of arbitrary dimension for our witness \eqref{QuadWit}.
By convexity, it follows that the quantity $W_N$ is maximized using a deterministic strategy \cite{gallego}. It then easily follows that

\be W_N \leq C_d = \frac{N(N-1)}{2} - \lfloor \frac{N}{d} \rfloor \left(N - \frac{d}{2} \left(\lfloor \frac{N}{d} \rfloor +1\right)\right) \ee
where $\lfloor x \rfloor $ is the integer part of $x$. Unless $N$ is a multiple of $d$, we have that $C_d<Q_d$, hence our witness can distinguish between classical and quantum systems of the same dimension. In particular, when $N$ is prime we have that $C_d<Q_d$ for any $d<N$. This is illustrated in Table~I, for the case $N=7$. Notably, dimension witnesses featuring such a 'quantum advantage' are potentially useful, for instance for the use of dimension witnesses in information-theoretic tasks \cite{marcin,hongwei}, and might be relevant to discuss problems in the foundations of quantum mechanics \cite{kleinmann}.

\begin{table}
	\begin{ruledtabular}
	\begin{tabular}{c||ccccccccc}
	$N$ &
	3 &
	4 &
	5 &
	6 &
	7 &
	8 &
	9 &
	10  \\
	\colrule
	$d$&  2   & 2,3 &   4 &   3,5 &   3,4,6 &   4,7 & 3,6,8 & 5,9 \\
	\end{tabular}
	\end{ruledtabular}
\caption{\label{tab:table2} The quantum dimension witness $V_N$ is tight for the above dimensions $d$; that is the bound $Q_d$ in \eqref{linearWit} can be attained.}
\end{table}

It turns out that our quadratic dimension witness can also be linearized, which might prove to be useful in certain situations. Here our main ingredient is the Cauchy-Schwartz (CS) inequality: for any real vector $\vec{v}$, we have that

\be
\sqrt{\sum_{i=1}^Mv_i^2}\geq \frac{1}{\sqrt{M}}\sum_{i=1}^M|v_i|,
\label{cauchy}
\ee

\noindent with equality iff $|v_i|=|v_j|$ for all $i,j=1,...,M$. Applying the CS inequality to our problem, i.e. to Eq. \eqref{QuadWit}, we obtain the linear dimension witness

\begin{eqnarray} \label{linearWit}
V_N &\equiv& \sum_{x>x'}P(x,(x,x'))-P(x',(x,x'))\\ \nonumber
& \leq& \frac{N\sqrt{N(N-1)}}{2}\sqrt{\left(1-\frac{1}{\min(d,N)}\right)}=Q_d.
\end{eqnarray}
Note that, the above inequality cannot always be saturated, hence the witness is not tight in general. However, one notable exception is the case $N=d+1$. 
Consider quantum states as given by eq.~(\ref{vectors}). Then, one can check that $|\braket{\Psi_x}{\Psi_{x'}}|=\frac{1}{d}$, for all $x\not=x'$. This means that, for optimal measurements, $ P(x,(x,x'))-P(x',(x,x'))=\sqrt{1-\frac{1}{d^2}} $
for all $x,x'$, and so the CS inequality (\ref{cauchy}) is saturated. Thus we finally get the tight linear dimension witness

\be
V_N  \leq Q_d = \frac{(d+1)\sqrt{d^2-1}}{2}.
\ee

Note that the witness \eqref{linearWit} turns out to be tight for other cases as well. For small values of $N$ and $d$, we investigated numerically the tightness of the witness (see Table~II), using the techniques of Ref. \cite{seesaw}. The case $N=d^2$ is of particular interest. Tightness is achieved for $N\leq 10$. Moreover, the optimal measurements are given here by symmetric informationally-complete (SIC) POVMs \cite{sic}. A study of the general case would be relevant, given the interest devoted to SIC POVMs.

Next we give the maximum value of $V_N$ for classical systems of dimension $d=N-1$, which is found to be $C_d=\frac{d(d+1)}{2}-1$. Since $C_d<Q_d$, this linear witness allows one to discriminate between quantum and classical systems, for all $d\geq 2$.

In conclusion, we have presented quantum dimension witnesses for systems of arbitrary Hilbert space dimension.
Notably our witnesses allow to discriminate between classical and quantum systems of the same dimension, a property that has already proven to be useful for information-theoretic tasks \cite{marcin,hongwei}.
Hence it would be interesting to investigate the robustness to noise and losses \cite{arno} of our witnesses.
Finally, our approach highlights a strong connection between dimension witnesses and quantum state discrimination, hence establishing a link between two research topics that have recently attracted attention, and might generate more results in the future.

\emph{Acknowledgements.} The authors acknowledge financial support from the UK EPSRC, the EU DIQIP, the Swiss National Science Foundation (grant PP00P2\_138917), the Hungarian National Research Fund OTKA (PD101461), and the Templeton Foundation.

\end{document}